\documentclass[11pt]{article}
\usepackage{color,rotating,pdflscape,colortbl}
\usepackage{graphicx}
\usepackage{amsmath}
\usepackage{amsthm}
\usepackage{amsfonts}
\usepackage{booktabs}
\usepackage{subfigure}
\usepackage{amssymb}

\newcommand{\Prob}{\mathrm{Pr}}
\newcommand{\NE}{\ell }
\newtheorem{theorem}{Theorem}[section]
\newtheorem{definition}[theorem]{Definition}
\newtheorem{lemma}[theorem]{Lemma}
\newtheorem{construction}[theorem]{Construction}

\newtheorem{corollary}[theorem]{Corollary}
\newtheorem{example}[theorem]{Example}

\newtheorem{remark}[theorem]{Remark}
\def\whitebox{{\hbox{\hskip 1pt
    \vrule height 6pt depth 1.5pt
    \lower 1.5pt\vbox to 7.5pt{\hrule width
            3.2pt\vfill\hrule width 3.2pt}%
    \vrule height 6pt depth 1.5pt
    \hskip 1pt } }}
\def\qed{\ifhmode\allowbreak\else\nobreak\fi\hfill\quad\nobreak
         \whitebox\medbreak}
\def\p{\noindent {\it Proof}: } 
\def\cg{\cellcolor{green}}   
\def\cy{\cellcolor{yellow}}

\begin{document}


\title{Combinatorial Analysis of Coded Caching Schemes
\thanks{R.\ Wei's research was supported by  
Natural Sciences and Engineering Research Council of Canada discovery grant RGPIN-05610.}
\author{Ruizhong Wei
\\
 Department of Computer Science\\
 Lakehead Univeristy\\
 Orillia, Ontario, L3V 0B9,  Canada\\
 \texttt{rwei@lakeheadu.ca}}}

\date{}
\maketitle

\begin{abstract}
Coded caching schemes are used to reduce computer network traffics in peak time. To determine the
efficiency of the schemes, \cite{MN} defined the information rate of the schemes and gave a construction
of optimal coded caching schemes. However, their construction needs to split the data into  a large number
of packets which may cause constraints in real applications. Many researchers then constructed new coded caching
schemes to reduce the number of packets but that increased the information rate. 
We define an optimization of coded caching
schemes under the limitation of the number of packets which may be used to verify the efficiency of 
these schemes. We also give some constructions for several infinite classes
of optimal coded caching schemes under the new definition.

\end{abstract}

\section{Introduction}

Predominantly driven by applications of internet such as audios, videos, big data etc,
there is a dramatic increase in network traffic now. The high
temporal variability of network traffic may cause   communication
systems to be congested during peak-traffic time and
underutilized during off-peak time.
Caching is an interesting strategy to cope with this high temporal
variability by shifting some transmissions from peak time to off-peak
time with the help of cache memories at the network edges.

To formally analyze the performance of the coded
caching approach,  Maddah-Ali and Niesen in \cite{MN} introduce a simple model of  coded caching schemes 
focusing on its basic structure.
In their setting,  $K$ users (nodes) are connected to
a server through a shared, error-free link. The server has a
database of $N$ files of equal size. Each of the users has access
to a cache memory which can store $M$ of the files. The scheme consists of two phases. 
\begin{enumerate}
\item  
The placement phase (during the off-peak time): each node's cache is filled with
partial data from the server. 
\item
The delivery phase, each user asks for
one of the $N$ files. The server then broadcasts some packets of data so that each
node can recover the requested file from the broadcasted information and the data in their cache.
\end{enumerate}

We assume that  no one knows which file a user will ask at the placement phase.
In the schemes proposed in \cite{MN}, each of the $N$ files in the server is divided into $F$ equal
sized packets. Each node  
stores $MF$ packets of files, where $0\le M \le N$, at the replacement phase.
 Since the node does not know which file it will be needed later, it will store equal size of packets for each of 
 the $N$ files. 
 At a later time, each of the nodes
requires a file (so $K$ files are requests, and we may assume these files are different). 
Then the server needs to broadcast data so that each
node can get the file it needs.  Two special cases are:
\begin{enumerate}
\item
$M = 0$: the server needs to broadcast $KF$ packets, since nothing is stored in nodes.
\item
$M = N$: the server does not need to broadcast packets, since the node has all of the files.
\end{enumerate}

In general, we need to consider  cases when $0< M < N$. In these cases, we want to know 
what is the minimum number of packets  the server
must broadcast, when $N, M,  K$ are given. In the model of \cite{MN},
the server will use the following strategy. It will xor some packets together
when it broadcasts packets.
If a node has all of the packets in the xored packet except one, then the node can recover the missing packet using its
stored packets. We use the following simple example from \cite{MN} 
to explain how to use this method to reduce the network traffic
during the the busy hours.

\begin{example}\label{e.demo}
{\rm
Suppose $N = 2, M = 1, F = 2, K = 2$. The two files  server has are $A$ and $B$, where $A$ is divided into packets $A_1$ and $A_2$,
and $B$ is divided into packets $B_1$ and $B_2$. In the placement phase, 
 node 1  stores $A_1, B_1$ and  node 2 stores $A_2, B_2$. Later, suppose  node 1 needs file $A$
and node 2 needs file $B$, then the server can broadcast $A_2 \oplus B_1$. Both nodes can recover the file
they needed.  It is not difficult to check that no matter which files the nodes
request, the server just needs to broadcast one packet to satisfy the nodes' requests.
}
\end{example}

\cite{MN} proposed an efficient scheme for general cases. They proved that their scheme is optimal in
terms of transmission rate, where the {\em transmission rate} is defined as the size of the packets required to
broadcast in the delivery phase if the values of $M, N$ and $K$ are given.
However, researchers noted   that their scheme requires to split each file into
$F$ packets, where $F$ increases exponentially with the number of nodes $K$. This will cause 
practical problems when $K$ is large. 
By trying to reduce the size of $F$ (which is called 
subpacketization), many researchers used various methods to construct new schemes (see, for examples \cite{CD, cjwy, K, sdlt, std, 
szg, tr, wcwc, wcwc1, YCTC, YTCC, ZCW}). Many combinatorial methods are used for these constructions including resolvable combinatorial designs,
linear block codes, $(6,3)$-free hypergraphs, project spaces, Ruzsa-Szem\'{e}redi graphs, strong edge coloring of bipartite graphs,
 Cartesion product, etc. 
Basically, these schemes have better subpacketizations but the transmission rate is not as good as that of
 the scheme in \cite{MN}. Non of these papers
proved that whether their schemes are optimal regarding to the efficiency under the size of their subpacketizations.
In fact, there is no clear criteria defined about the optimization of those schemes, if we need to limit the size of 
subpacketization. 

In this paper,
we define the optimization of a coded caching scheme
under the conditions that the size of subpacketization, the size of database, the number of nodes and the size of 
storage of the nodes are given. 
This definition is ``weaker'' than the definition of transmission rate. However,
under the practical constraints, this will provide a benchmark for various coded caching schemes. 

Yan et al in \cite{YCTC} introduced an elegant notion to describe a coded cache scheme called placement delivery array (PDA). 
It turns out that most constructions of the coded caching schemes mentioned above can be described as some PDAs. 

In this paper, we use combinatorial method to analysis the coded caching schemes that can be described using PDAs,
because PDA is an combinatorial object in nature. We will focus
on the number of packets the server needs to deliver during the peak-traffic time for any given number of the nodes, any size of
the files and any storage size available on nodes.
So we want to determine what is the smallest size of $|S|$, if $N, M, K, F$ are given. 
Comparing to the transmission rate, we want to add the parameter $F$ to the definition
of optimization.
 Following  the setting of \cite{MN}, however, we assume that all of
the files stored in
the server have
the same size and all of the nodes in the system have the same size of storage. 

The rest of this paper are arranged as follows. Section \ref{s.PDA} describes the PDA and defines
the optimization of PDAs. Some simple examples of optimal PDAs are  given in
this section. Section \ref{s.RPDA} discusses a special class of optimal PDA called RPDA. A general
construction of the RPDAs are given. Section \ref{s.s} gives several classes of optimal PDAs and some
examples for special parameters.
Section \ref{s.conclusion} summarizes the results of this paper and suggests some open research
problems.

\section{Placement delivery array}\label{s.PDA}
 
In this section, we introduce the placement delivery array.  \cite{YCTC} proposed an interesting and 
simple combinatorial structure, 
called {\em placement delivery array (PDA)}. 
Most coded caching schemes constructed by many researchers can be described as building  PDAs.

 We give the definition of the PDA below. Our definition is slightly different from the description of \cite{YCTC}, but 
 essentially they are  the same.

\begin{definition}\rm
\label{def-PDA}
A {\em placement delivery array (PDA)} is an $F\times K$ array ${R}=(r_{j,k})$, $1\leq j \leq F, 1 \leq k\leq K$, 
defined on an $s$-set $S$
 such that 
\begin{enumerate}
  \item  each cell of the array is either empty or an element of $S$,
  \item each column contains $Z$ empty cells,
\item
each row and each column contains an element $t \in S$ at most once.
  \item  if any two nonempty cells $r_{j_1,k_1}$ and $r_{j_2,k_2}$ are the same,   i.e., $r_{j_1,k_1}=r_{j_2,k_2}= t, t \in S $,  then
      $r_{j_1,k_2}$ and $r_{j_2,k_1}$ are empty.
   \end{enumerate}
\end{definition}

In what follows, we will use the notation $S$-PDA$( F, K, Z)$ to denote the design and  set $S = \{ 1, 2, \cdots, s\}$
in most cases. We
also use $0$ to denote the empty cells in some occasions for convenience.

The parameters in a PDA corresponding to a coded caching scheme are as follows.
\begin{enumerate}
\item
$F$ is the number of packets for each of the files. The server has total $FN$ packets.
\item
$K$ is the number of users (nodes).
\item
$Z$ represents the size of node storage. A node can store $Z$ packets of each file. So the node can store
$ZN$ packets. 
\item
$s$ is the number of packets which  the server needs to deliver  at the peak-traffic time. 
\end{enumerate}

In a coded caching scheme, each of the $N$ files in the server is divided into $F$ packets, indexed
as $1, \dots, F$. 
 The $i$th row of the $PDA$ correspondents to the $i$th packets of 
the $N$ files. Each column of a $PDA$ correspondents to a user. If $r_{i,j}$ is empty, then the node $j$
will store the $i$th packet of each of the $N$ files. When the $K$ users request the files, say files $f_1, \dots,
f_K$, the server xor's the packets according to the elements of $S$. For example, if $r_{i_1, j_1} = 
r_{i_2, j_2} = \dots = r_{i_u, j_u} = t \in S$, then the $i_1$th packet of  the $f_{j_1}$, 
the $i_2$th packet of  the $f_{j_2}, \dots,$ the $i_u$th packet of  the $f_{j_u}$ are xored together. After
the server broadcasts the $s$ packets, each user will be able to recover the file they asked by using
the data in their caches, because from 4 of 
Definition \ref{def-PDA} the node has all of the other packets excepting the packet of their requested file
in the xored packets.

\begin{example}
\rm
The corresponding PDA of the scheme in Example \ref{e.demo} is as follows.

\[
\begin{array}{|c|c|}
\hline
&1\\
\hline
1&\\
\hline
\end{array}
\]
\end{example}

From the above description we can see that the efficiency of the coded caching scheme depends on
the size of $|S|$ if the values of $F, K$ and $Z$ are given.

Therefore we give the following definition.

\begin{definition}\rm
Let $s(F, K, Z) = \min \{ |S| : \mbox{ there is an $S$-PDA$(F, K,Z)$}\}$. Then we say that an 
$S$-PDA$(F, K,Z)$  is optimal if $|S| = s (F, K, Z)$.
\end{definition}

We first look at some simple cases and prove the following.

\begin{lemma}\label{l.simple}
For some  special values of $Z$, we have
\begin{enumerate}
\item $s( F, K, 0) = FK$;
\item $s(F, K, F) = 0$;
\item $s (F, K, F - 1) = \lceil {K\over F}\rceil$;
\item If $F \ge tK, $ then $s(F, K, F-t) = t$.
\end{enumerate}
\end{lemma}

\p The proof of 1 and 2 are straightforward.

3. Since each element of $S$ only can appear in at most one cell in a row or a column, we have  $s (F, K, F - 1) \ge \lceil {K\over F}\rceil$.
For an $F\times K$ array, we divide it into separated $F\times F$ sub-arrays (one of them may have less than $F$ columns) and give the main 
diagonal of each sub-array an element of $S$, which forms the required PDA.

4. For any $S$-$PDA( F, K, F- t)$, $|S| \ge t$, since each column
has   $t$ different elements.
 For an $F\times K$ array, let the columns be $[ 1, 2, \cdots t, 0, \cdots, 0]^T, 
[\underbrace{0, \cdots, 0}_t, 1,2, \cdots, t, 0, \cdots, 0]^T, \linebreak
\cdots, [ \underbrace{0, \cdots, 0}_{t(K - 1)}, 1, 2, \cdots, t, \cdots]^T$, which forms the required PDA (0 denotes the empty cell).
\qed

The following examples are from  3 and 4 of Lemma \ref{l.simple}.

\begin{example}\label{e.2-3}
An optimal 3-$PDA(4, 10, 3)$ and a optimal 2-$PDA(8,4, 6)$:

\hskip 2 cm
$
\begin{array}{|c|c|c|c|c|c|c|c|c|c|}
\hline
1&&&&2&&&&3&\\
\hline
&1&&&&2&&&&3\\
\hline
&&1&&&&2&&&\\
\hline
&&&1&&&&2&&\\
\hline
\end{array}
$
\hskip 1 cm
$
\begin{array}{|c|c|c|c|}
\hline
1&&&\\
\hline
2&&&\\
\hline
&1&&\\
\hline
&2&&\\
\hline
&&1&\\
\hline
&&2&\\
\hline
&&&1\\
\hline
&&&2\\
\hline
\end{array}
$
\end{example}

In what follows, we always assume that $F \ge Z + 1$.

In the following proofs, we will use the fact that for an $S$-PDA$(F, K, Z)$,
 if we exchange rows, or exchange columns, or exchange  elements, it is still an $S$-PDA$(F, K, Z)$.

Let $u = F - Z$. Then we can always let the first column of the $S$-PDA$(F, K, Z)$ has 
the form $[ 1, 2, \cdots, u, 0, \cdots, 0]^T$ by exchange rows and elements.
We call this form of a PDA a {\em regular PDA}.

\begin{lemma}\label{l.upsub}
 In a regular $S$-$PDA( F, K, Z)$ with $u = F - Z$, the columns of the upper $u \times K$ sub-array of the PDA contain no
element in $\{ 1, 2, \dots, u \}$ except the first column.
\end{lemma}

\p If there is a $t \in \{ 1, 2, \dots, u\}$ in a column other than the first column, then the condition 4 of Definition \ref{def-PDA} will be 
violated, because the first column of the sub-array contains no empty cell. 
\qed

\begin{lemma}\label{l.k=2}
 $$s(F, 2, Z) = \left\{
\begin{array}{ll}
2F - 3Z&\mbox{\rm if $F \ge 2Z$}\\
F -Z &\mbox{ \rm otherwise}
\end{array}\right.$$
\end{lemma}
\p In a regular $S$-$PDA( F, 2, Z)$, the second column has at most $Z$ elements same as that in first column.
Therefore first column contains $F-Z$ different elements and second column contains at least $F - 2Z$ elements
which are different from the elements in first column.  On the other hand, it is easy to construct $S$-$PDA( F, 2, Z)$ with the 
minimum size of $S$.
\qed

To optimize a PDA, we want each element to appear in the PDA as many times as possible to reduce the size of $S$.
On the other hand, the properties of PDA will limit the frequency of the elements. The following lemma gives one of
the  important 
limitations of the frequency. 

\begin{lemma}
In a $PDA(F,K,Z)$, one element  can appear at most $Z + 1$ times. 
\end{lemma}
\p Suppose $r_{g,i} = t \in S$.   If there is another cell $r_{h,j} =t$, then $i\neq j$ and $g \neq h$ by the definition of PDA.
It also follows from the definition of PDA that $r_{h, i}$ must be empty. Since each column has $Z$ empty cells, there are at most
$Z$ different values of $h$. Therefore
element $t$ appears at most $Z$ times in other columns.
\qed

From the above lemma, we have the following corollary.

\begin{corollary}
An $PDA(F,K,Z)$ is optimal, if each element  of $S$ appears $Z + 1$ times in the array. 
\end{corollary}

We have one lower bound of $s(F, K, Z)$.

\begin{theorem}\label{t.lowbound}
In any $S$-$PDA(F, K, Z)$, 
\begin{equation}
|S| \ge \left\lceil { K(F- Z)\over Z + 1}\right\rceil.  \label{n.bound1}
\end{equation}
\end{theorem}

\p There are $FK$ cells in a PDA$(F, K, Z)$ and $KZ$ of them are empty. Since each element appears 
at most $Z+ 1$ times, $|S| \ge {FK -KZ \over Z+ 1}$ .
\qed

 For $Z = 1$, we can easily obtain the following result.

\begin{lemma}\label{l.z=1}
$s(F,F, 1) = {F(F-1)\over 2}$.
\end{lemma}
\p We construct the required PDA as follows. Let the main diagonal of the $F\times F$ array contains empty cells.
For the rest of upright half array, fill first row with $1, 2, \dots, F - 1,$ fill second row with $F, F+1, \dots, 2F -3$
$\dots \dots$, and fill the $(F - 1)$th row with $F(F-1)\over 2$. The required array can be obtain by filling the other
cells symmetrically. Since each element appears twice in the array, it is optimal from Theorem \ref{t.lowbound}.
\qed

\begin{example}\label{e.2}
The following $4$-$PDA(4, 6,2)$ is optimal since each element appear 3 times:

$$
\begin{array}{|c|c|c|c|c|c|}
\hline
&&&4&3&2\\
\hline
&4&3&&&1\\
\hline
4&&2&&1&\\
\hline
3&2&&1&&\\
\hline
\end{array}
$$
\end{example}

\section{RPDA and its constructions}\label{s.RPDA}

Lemma \ref{l.z=1} and the cases 1, 2 and 3 of Lemma \ref{l.simple}  show that the bound in Theorem \ref{t.lowbound} is exact
in many special cases.  On the other hand, the case 4 of Lemma \ref{l.simple} and Lemma \ref{l.k=2} show that the bound in 
Theorem \ref{t.lowbound} cannot be
reached in general. It is an interesting question that for what parameters of $F, K, Z$, the low bound of Theorem \ref{t.lowbound}
can be reached. So we give the following definition.

\begin{definition}\rm
\label{def-RPDA}
A {\em restricted placement delivery array (RPDA)} is an $F\times K$ array
 ${R}=(r_{j,k})$, $1 \leq j \le F, 1 \leq k \le K$, defined on a set $S$
  that satisfies all of the conditions 1 - 4 of Definition \ref{def-PDA}
 plus 
 \begin{enumerate}
 \item[5.]
 $ |S| = \left\lceil { K(F - Z)\over Z + 1}\right\rceil. $
 \end{enumerate}
\end{definition}

An RPDA is an optimal PDA, but an optimal PDA may not be an RPDA.

\begin{lemma}\label{l.necessary}
 One necessary 
condition for the existence of RPDA is 
\begin{equation}
K >  {(Z + 1)(F-Z -1)\over F-Z}.
\end{equation}
\end{lemma}
\p
If $K \le  {(Z + 1)(F-Z -1)\over F-Z}$, then ${K(F - Z)\over Z+1} \le {(F-Z-1)}$. 
But in any $S$-PDA, $|S| \ge F - Z$.
\qed

Niu and Cao gave an upper bound for $s(F, K, Z)$ in \cite{NC} as follows.

\begin{lemma}\cite{NC}\label{l.boundNC}
Suppose there is an $|S|$-PDA$(F, K, Z)$, then 
\begin{eqnarray}
|S| &\ge &\left\lceil {(F-Z)K \over F}\right\rceil + \left\lceil {F-Z - 1\over F-1}\left\lceil {(F-Z)K \over F}\right\rceil\right\rceil 
+ \cdots \nonumber \\ 
&&+ \left\lceil {1\over Z+1} \left\lceil {2\over Z+ 2}\left\lceil \cdots \left\lceil {(F-Z)K \over F}\right\rceil \cdots  
\right\rceil\right\rceil\right\rceil. \label{n.bound2}
\end{eqnarray}
\end{lemma}

From Lemma \ref{l.boundNC}, we have the following.

\begin{theorem}\label{t.rpda}
If there is an $|S|$-PDA$(F, K, Z)$ 
with $|S| =   {K(F - Z) \over Z+1} $.
Then $K = \ell{F\choose Z}$ for some integer $\ell \ge 1$.
\end{theorem}
\p From Lemma \ref{l.boundNC} we have
\begin{eqnarray}
|S| &\ge&\left\lceil {(F-Z)K \over F}\right\rceil + \left\lceil {F-Z - 1\over F-1} {(F-Z)K \over F}\right\rceil
+ \cdots 
+ \left\lceil {1\over Z+1}  {2\over Z+ 2}   \cdots {(F-Z)K \over F}  \right\rceil  \nonumber \\
 &\ge&   {(F-Z)K \over F}  +   {F-Z - 1\over F-1} {(F-Z)K \over F} 
+ \cdots 
+   {1\over Z+1}  {2\over Z+ 2}   \cdots {(F-Z)K \over F} \label{n.thrpda}\\
&=& \left({(F-Z)\over F}  +   {(F-Z - 1)(F-Z) \over (F-1)F} 
+ \cdots 
+   {1\cdot 2 \cdots (F-Z)\over (Z+1)( Z+ 2)   \cdots F}\right) K \nonumber \\
&=& \left({(F-Z)\over F}{F\choose Z}  +   {(F-Z - 1)(F-Z) \over (F-1)F} {F\choose Z}
+ \cdots 
+   {1 }\right) {K \over {F\choose Z}} \nonumber \\
&=& \left( {F-1 \choose Z}+ {F-2 \choose Z} + \cdots + {Z \choose Z}\right){K \over {F\choose Z}} \nonumber \\
&=& {F\choose Z+1}K\over {F\choose Z} \nonumber \\
&=& {(F -Z)K\over Z + 1}. \nonumber
\end{eqnarray}

 Note that (\ref{n.bound2}) is equal to (\ref{n.thrpda}), only if $ {(F-Z)K \over F}, {F-Z - 1\over F-1} {(F-Z)K \over F} ,\cdots 
, {1\over Z+1}  {2\over Z+ 2}   \cdots {(F-Z)K \over F}$ are all integers.
From the last term of (2), when $(1) = (2)$, ${ (F - Z)! Z!\over F!}K$ is an integer.  Therefore 
$K = \ell {F \choose Z}$.  
 \qed
 
 Therefore we first consider the existence of PDA$(F, {F\choose Z}, Z)$.
 
From Lemmas \ref{l.simple} and  \ref{l.z=1}, we have the following result.

\begin{theorem}
There exist an RPDA$(F,F,1)$ and RPDA$(F, K, F -1)$ for any integer $F\ge 2$.
\end{theorem}

When $K(F- Z)\over Z + 1$ is an integer, then we have the following result.

\begin{lemma}\label{l.times}
Suppose $K(F- Z)\over Z + 1$ is an integer and there is an RPDA$ (F, K, Z)$. Then there is an 
RPDA$( F, \ell K, Z) $ for any positive integer $\ell$.
\end{lemma}
\p
 Make $\ell$ copies of the RPDA$(F, K, Z)$, using different elements for each of the copy. So there are
 $\ell K(F- Z)\over Z + 1$ elements used. Put these copies together to make an $F \times \ell K$
 array. It is readily checked that the array is an RPDA$( F, \ell K, Z) $.
 \qed

When $K(F- Z)\over Z + 1$ is not an integer, then the above construction does not work, because in general
$\ell \left\lceil {K(F- Z)\over Z + 1}\right\rceil  >\left\lceil {\ell K(F- Z)\over Z + 1}\right\rceil$.

\begin{lemma}\label{l.minus}
Suppose $K(F- Z)\over Z + 1$ is an integer and there is an RPDA$ (F, K, Z)$. Then there is an 
RPDA$( F, K - x, Z) $ for  $0 \le x < {Z + 1\over F - Z}$.
\end{lemma}
\p
 Since $0 \le x < {Z + 1\over F - Z}$, we have
\begin{align*}
{(K - x)(F- Z)\over Z + 1}& >{ ( K - {Z + 1\over F - Z})(F-Z) \over Z+1}\\
& = {K (F - Z)\over Z + 1} -1.
\end{align*}
Therefore $\left\lceil {(K - x)(F- Z)\over Z + 1}\right\rceil = {K(F- Z)\over Z + 1}$. 
If we delete $x$ columns from a RPDA, it is still a PDA. When $x < {Z + 1\over F - Z}$,
the result PDA is an RPDA.
\qed

The PDA in Example \ref{e.2} is an RPDA$( 4, 6, 2)$. By the above lemma, we have a
RPDA$(4, 5,2)$. 


\begin{lemma}\label{l.462}
There exist RPDA$(4, 6\ell, 2)$ and RPDA$(4, 6\ell -1, 2)$ for all $\ell \ge 1$.
\end{lemma}
\p
The conclusion follows from Example \ref{e.2} and Lemmas \ref{l.times} and \ref{l.minus}.
\qed

In an RPDA, each element appears at the same number of cells. Therefore the array has some ``balance''
property. We consider the cases that the empty cells distribute evenly in the array.
It turns out that these cases  can  reach  the bound (\ref{bound}) in Theorem \ref{t.lowbound}.

 Next,  we give
a detailed  recursive construction of infinite RPDAs. We start from the case $Z =1$. Although Lemma \ref{l.z=1}
gives the proof of the existence of RPDA$(F, F, 1)$, we will give the following recursive construction
to fit in our general recursive method.

\begin{construction}\label{c.1}
There exists an RPDA$(F, {F\choose 1}, 1)$ for any positive integer $F \ge 2$.
\end{construction}
\p For $F= 2$, we construct a $2\times 2$ array $A = (a_{i,j})$ such that $a_{1, 1}$ and $a_{2, 2}$
are empty and other two cells contains a same element 1.  Suppose we already have constructed 
RPDA$(k,k,1)$. Then we construct a RPDA$(k+1, k+1, 1)$ by add one column in the left of $A$ 
and one row on the top of the resulting array such that the elements on the top row from the last column 
to the second column are ${k(k-1)\over 2} +1, \dots, {k(k-1)\over 2} +k$ and the first cell is empty. 
The first column from the bottom
to top except the first empty cell are ${k(k-1)\over 2} +1,\cdots, {k(k-1)\over 2} +k$.  
\qed

\begin{remark}\rm
We can view the first column except the first cell at the above construction as an  RPDA$(k, 1,0)$. 
This idea is used in our general constructions. Also note that we used an ``inverse order''
 for $S$ in the resulting RPDA.
\end{remark}

\begin{example}\label{e.1}
The following RPDA$(5, 5,1)$ is constructed using Construction \ref{c.1}.

$$
\begin{array}{|c|c|c|c|c|}
\hline
&10&9&8&7\\
\hline
10&&6&5&4\\
\hline
9&6&&3&2\\
\hline
8&5&3&&1\\
\hline
7&4&2&1&\\
\hline
\end{array}
$$
\end{example}

Next we consider $Z = 2$.

\begin{construction}\label{c.2}
There exists an RPDA$(F, {F\choose 2}, 2)$ for any positive integer $F \ge 3$.
\end{construction}
\p
An $3\times 3$ array with 1's at its anti-diagonal
and empty for other cells  is an RPDA $(3,{3\choose 2},2)$.  
 Suppose
we already constructed an RPDA$(k, {k\choose 2}, 2)$ using the recursive method.
 We can construct an RPDA$(k + 1, {k+1 \choose 2}, 2)$
as follows. 

Let $A$ be a RPDA$(k, {k\choose 1}, 1)$ constructed from Construction \ref{c.1},
and $B$ be a RPDA$(k, {k \choose 2},\linebreak 2)$. But the elements in $B$ are
$\{ 1, 2, \dots, {k\choose 3} \}$ and the elements in $A$ are $\{ {k\choose 3} +1, 
{k\choose 3} + 2, \dots , {k+ 1\choose 3} \}$. 
Similar to   Construction \ref{c.1}, we put $AB$ together and add one row on the top of the
resulting array to form the array $R$. The first $k$ cells of the added row are empty. The other cells in  
this row are ${k+ 1\choose 3}, {k+ 1\choose 3} -1, \dots, {k\choose 3} + 1$, from the $(k+1)$th cell to the
end of the row 
respectively. 

It is easy to check that the conditions 1, 2 and 3 of Definition \ref{def-PDA} are satisfied. For condition 4, we just need
to check the elements at the  left part of the array (the elements in $A$). These elements  also appear in the first 
row above  the
array $B$.
Since in each recursion of RPDA$(k, {k \choose 2}, 2)$ we add the empty cells in the top row of 
RPDA$(k - 1, {k - 1\choose 1}, 1)$ and 
the main diagonal of the sub-array, the  empty cells in $B$ are as follows.
 The  two empty cells in first column of $B$ are row 1 and row 2, which we denoted as $(1, 2)$.
 Then two empty cells in other columns of $B$ are $(1,3),\cdots, (1, k), (2, 3),(2, 4), \cdots, (2,k), \cdots, (k-1, k)$,
 respectively. So the empty cells in the right part of $R$ are $(2,3),\cdots, (2, k), (3, 4),(3, 5), \cdots, (3,k+1), \cdots, (k, k+1)$,
 respectively.
 Since the two row positions of an element  $t \in A$ is exactly  the positions of empty cells of the 
column   where $t$ is on the top of $ B$, and the top row over $A$ contains empty cells, 
the condition 4 is satisfied.
\qed

\begin{remark}\rm\label{r.2}
In an RPDA$(F, {F\choose 2}, 2)$ constructed by Construction \ref{c.2}, $S = \{ 1, 2, \dots, {F\choose 3}\}$.
The row positions of elements of $S$ are as follows.

\[
\begin{array}{c|ccc}
\hline
elements&&rows&\\
\hline
1&F-2&F-1&F\\
2&F-3&F-1&F\\
3&F-3&F-2&F\\
4&F-3&F-2&F-1\\
5&F-4&F-1&F-2\\
\cdots&\cdots\\
10&F-4&F-3&F-2\\
\cdots&\cdots\\
{F\choose 3}&1&2&3\\
\hline
\end{array}
\]
\end{remark}

\begin{example}\label{e.5-10-2}
The following example is an RPDA$(5,10,2)$, which is obtained by applying the Construction \ref{c.2}
twice.

$$
\begin{array}{|c|c|c|c|c|c|c|c|c|c|}
\hline
&&&&10&9&8&7&6&5\\
\hline
&10&9&8&\cg&\cg&\cg &\cg 4&\cg 3&\cg 2\\
\hline
10&&7&6&\cg &\cg 4&\cg 3&\cy&\cy&\cy 1\\
\hline
9&7&&5&\cg 4&\cg&\cg 2&\cy&\cy 1&\cy\\
\hline
8&6&5&&\cg 3&\cg 2&\cg &\cy 1&\cy &\cy \\
\hline
\end{array}
$$
\end{example}

To further explain our general recursive construction,  we construct some infinite class of RPDA with $Z = 3$
before we describe our general construction.

\begin{construction}\label{c.3}
There exists an RPDA$(F, {F\choose 3}, 3)$ for any integer $F \ge 4$.
\end{construction}
\p We use a recursive construction similar to the Construction \ref{c.2}. We start from a RPDA$( 4, 4, 3)$
which is a $4\times 4$ array with  all empty cells except  the cells of its anti-diagonal that are filled with
element 1.  Then we recursively use the following
method. Suppose we already have a RPDA$(k, {k\choose 3}, 3)$ created using this method. 
To construct an RPDA$(k+1, {k+1\choose 3},3)$, we
first put an RPDA$(k, {k\choose 2}, 2)$, $A$, which is constructed by Construction \ref{c.2}
 and an RPDA$(k, {k\choose 3}, 3)$, $B$,  together to form a $k \times {k+1\choose 3}$ array
(note that ${k\choose 2} + {k \choose 3} = {k+1\choose 3}$). 
The elements used in $B$ are $\{ 1, 2, \dots, {k\choose 4}\}$ and elements in $A$ are $\{ {k\choose 4} + 1,
{k\choose 4} + 2, \dots, {k\choose 4} +  {k\choose 3}\}$. 
(Note that ${k\choose 4} +  {k\choose 3} = {k+1\choose 4}$).
 Add one row at the top of the array with
${k\choose 2}$  empty cells  followed by  ${k \choose 3}$ elements of $A$ in the inverse order. (In an
RPDA$(k, {k\choose 2}, 2), |S| = {k\choose 3}$). Similar to the Construction \ref{c.2}, we basically 
need to prove that the condition 4 of Definition \ref{def-PDA} is satisfied.

From the recursive construction, we can see that the empty cells in $B$ from the first column to the 
last column are $(1, 2,  3), (1, 2, 4), \cdots, (1, 2, k), (1, 3, 4), \dots, (1, 3, k), \dots, (2,3,4),\linebreak
\cdots, (2, 3, k), (2, 4,k), \dots,  \dots, (k-2, k-1, k)$. From the Remark \ref{r.2}, condition 4 of the 
Definition \ref{def-PDA} is satisfied.
\qed

\begin{example}\label{e.6-20-3}\rm
The   example in Figure \ref{f.z=3} is an RPDA$(6,20,3)$ from the construction \ref{c.3},
which contains an RPDA$(5,10,3)$ and an RPDA$ (4,4,3)$.
\begin{figure}
$$
\begin{array}{|c|c|c|c|c|c|c|c|c|c|c|c|c|c|c|c|c|c|c|c|}
\hline
&&&&&&&&&&15&14&13&12&11&10&9&8&7&6\\
\hline
&&&&15&14&13&12&11&10&\cg&\cg&\cg&\cg&\cg&\cg&\cg 5&\cg 4&\cg 3&\cg 2\\
\hline
&15&14&13&&&&9&8&7& \cg&\cg&\cg&\cg 5&\cg 4&\cg 3&\cy&\cy&\cy&\cy 1\\
\hline
15&&12&11&&9&8&&&6&\cg&\cg 5&\cg 4&\cg&\cg&\cg 2&\cy&\cy&\cy 1&\cy\\
\hline
14&12&&10&9&&7&&6&&\cg5&\cg&\cg 3&\cg&\cg 2&\cg &\cy&\cy 1&\cy &\cy \\
\hline
13&11&10&&8&7&&6&&&\cg 4&\cg 3&\cg&\cg 2&\cg&\cg&\cy 1&\cy&\cy&\cy\\
\hline
\end{array}
$$
\caption{An RPDA$(6,20,3)$}\label{f.z=3}
\end{figure}
\end{example}

Based on Constructions \ref{c.1}, \ref{c.2} and \ref{c.3}, we will prove a general construction.
For convenience, we define some order for all the combinations of $r$ elements from the set
$\{ 1, 2, \dots, n\}$ as follows. 

$
\{ 1, 2, \dots, r\}, \{ 1, 2, \dots, r-1, r+1\}, \dots, \{1, 2, \dots, r-1, n\}, \{1, 3, \dots, r+1\}, \dots, 
\{1, n-r+2, \dots, n\}, $ 
$\{ 2, 3, \dots, r+1\}, \dots, \{ 2, n-r+2, \dots, n\}, \dots, \dots, \{n-r+1, \dots, n\}.
$
   
\begin{figure}[h]
 \includegraphics[width=\textwidth]{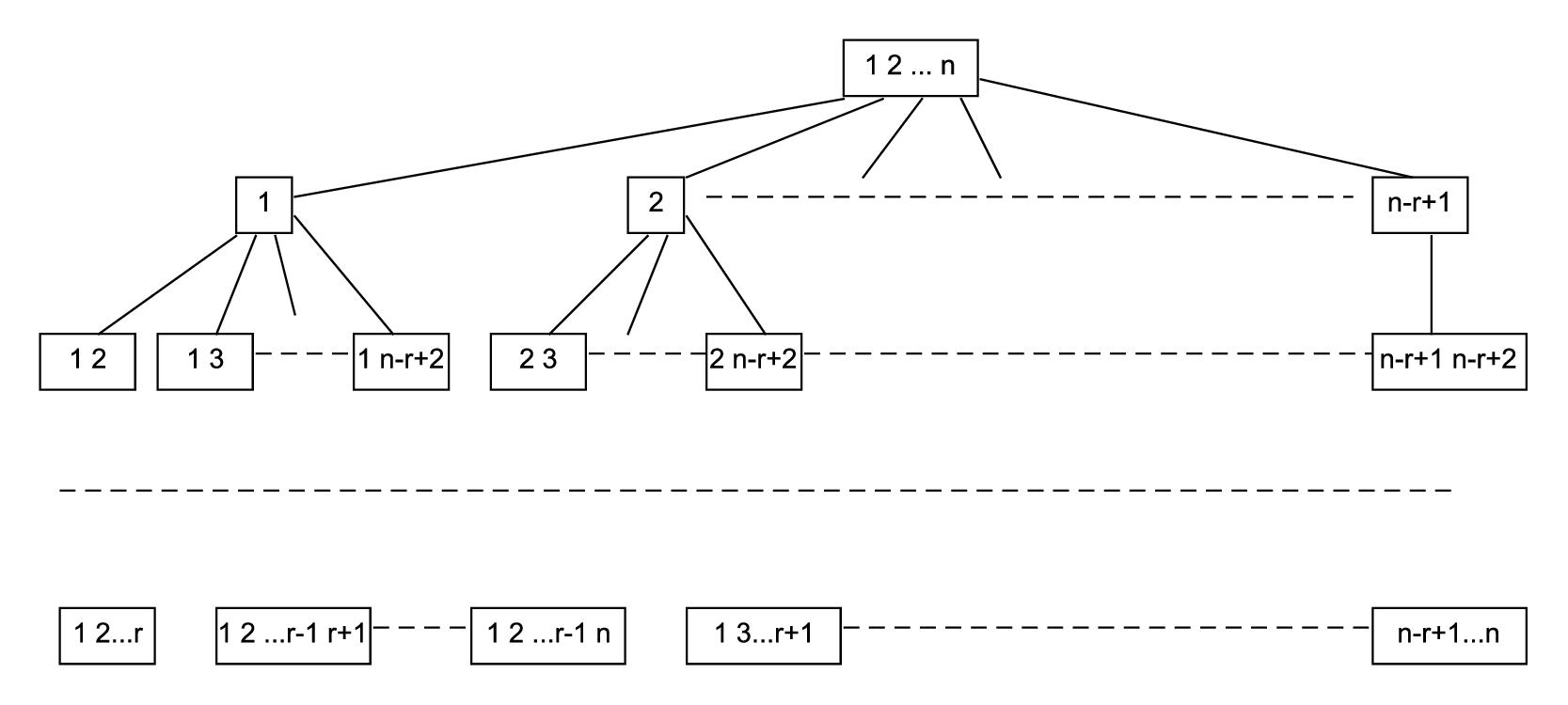}
 \caption{TB order of combination}\label{f.combi}
 \end{figure}

 We call the above order as {\em TB order } (top-bottom order).
 The Figure \ref{f.combi} shows how to create the combinations in TB order.
 
 When the combinations are created from the inverse set $\{ n, n-1, \dots, 1\}$ instead of 
the set  $\{ 1, 2, \dots, n\}$, we say that they are in  {\em BT order}. 
 
\begin{theorem}\label{t.general}
There exists an RPDA$(F, {F\choose Z}, Z)$ for any integer $F \ge Z+1$.
\end{theorem}
\p
We are going to prove that  for any given $Z > 1$ and $k \ge Z+1$, we have an RPDA$(k, {k\choose Z}, Z)$ which satisfies the 
following combination properties.
\begin{enumerate}
\item
The empty
cells of the RPDA from column 1 to column ${k\choose Z}$ are in TB order. So the empty cells in column 1 are
at positions in $\{1, 2, \dots, Z\}$, the empty cells in column 2 are at positions in $\{ 1, 2, \dots, Z-1, Z+1\}$, $\dots$,
 the empty cells
in column $k\choose Z$ are at positions in $\{k-Z+1, k-Z +2, \dots k\}$. 
\item
The rows where the numbers of
$S$ appears are in BT order. Therefore the number 1 appears in rows $\{ k -Z -1, \dots, 
k \}$, the number 2 appears in rows $\{ k  -Z -1, k  -Z + 1\dots, 
k \}$, $\dots,$ the number $\frac {{k\choose Z} (k - Z )}{Z+1}$ appears in rows
$\{1, 2, \dots, Z+1\}$. 
\end{enumerate}

Constructions \ref{c.2} and \ref{c.3} showed that when $Z = 2$ and $3$, the above described RPDA exist.

To construct an RPDA$(k+1, {k+1\choose Z}, Z)$ for any integer $k + 1 \ge Z+1$, we start with an RPDA$(Z+1, {Z+1\choose Z}, Z)$
which is a $(Z+1)\times (Z+1)$ array with 1's at its cells of anti-diagonal and empty for other cells. Suppose we already have 
 an RPDA$(k , {k \choose Z -1}, Z -1)$, $A$, and an RPDA$(k, {k\choose Z}, Z)$, $B$.  The $S$ set of $B$ contains $\left\{ 1, 2, \dots, 
 \frac{{k \choose Z} (k- Z)}{Z+ 1}\right\}$ and the set in $A$ contains $\left\{   
\frac{{k \choose Z} (k- Z)}{Z+ 1} + 1 , \dots, \frac{{k \choose Z} (k- Z)}{Z+ 1}+ \frac{{k \choose Z -1}(k - Z+ 1)}{Z}\right\}$.

Note that 
\begin{align*}
&\frac{{k \choose Z} (k- Z)}{Z+ 1}+ \frac{{k \choose Z -1}(k - Z+ 1)}{Z}\\
&=\frac{1}{Z(Z+1)}\left({k \choose Z} (k- Z)Z + {k \choose Z -1}(k - Z+ 1)(Z+1)\right)\\
&=\frac{1}{Z(Z+1)}\left({k \choose Z} (k- Z)Z + {k \choose Z -1}((k - Z)Z+ k+1)\right)\\
&= \frac{1}{Z(Z+1)}\left({k + 1\choose Z} (k- Z)Z + {k \choose Z -1}( k+1)\right)\\
&= \frac {{k + 1\choose Z} (k- Z)}{Z+1} + \frac{{k \choose Z -1}(k+1)}{Z(Z+1)}\\
&= \frac {{k + 1\choose Z} (k + 1- Z )}{Z+1}. 
\end{align*}
 
We  construct a $(k+1) \times {k+1\choose Z}$ array as follows. We add one row on the top of the arrays $AB$
to form an array $R$. In the new row (the first row of $R$), the first $k \choose Z-1$ columns are empty and
the the other columns of the first row are filled with the elements $\left\{   
\frac{{k \choose Z} (k- Z)}{Z+ 1} + 1 , \dots, \frac{{k \choose Z} (k- Z)}{Z+ 1}+ \frac{{k \choose Z -1}(k - Z+ 1)}{Z}\right\}$
in the inverse order. Now we can check that $R$ satisfies the conditions of Definition \ref{def-PDA}.

Since $A$ and $B$ are RPDAs with different elements, and we put empty cells on the top of $A$ and 
different elements of $A$ on the top of $B$,
the conditions  1 - 3 of Definition \ref{def-PDA} are satisfied. All of the elements appears in $B$  satisfy the condition 4 of Definition
\ref{def-PDA}. For elements appear in $A$, we just need to verify that an element appears in $A$ and the same 
element in the row above
$B$ satisfy the condition 4, since the same element both appear in $A$ already satisfies the condition 4.
 Suppose $r_{1, t_1} = r_{s, t_2}$ is the
same element, where $r_{1, t_1}\in B$ and $r_{s, t_2} \in A$,
i.e., $t_1 > {k\choose Z -1}, s > 1$ and $t_2 \le {k\choose Z-1}$. Then $r_{1, t_2}$ is empty by the construction. 
To show that $r_{s, t_1}$ is also empty, we note that the rows of one element in $A$ are arranged in BT order (each of them appears $Z$ 
times in $A$ and $Z + 1$ times in $R$). The empty cells in $B$ are in TB order, but we have arranged the elements
in the row above $B$  in the inverse order. Therefore
$r_{s, t_1}$ must be an empty cell. This proves that $R$ is an RPDA. 

Finally, we need to prove that $R$ also satisfies the two  combination properties. Since we added a new row on the top to form $R$,
the original row numbers of $A$ and $B$ will increase one. The first $k \choose Z-1$ columns have empty row one and other
empty cells are in the TB order of combinations of $Z -1$ from $\{ 2, \dots, k+1\}$. The  empty cells at remaining columns are 
the TB order of combinations 
of $Z$ from $\{ 2, \dots, k+1\}$. Therefore the empty cells in $R$ is in the TB order of combinations of $Z$ from $\{1, \dots,
k+1\}$.  

 The number of rows where the elements appear in $A$ or $B$ are in BT orders.  Now we have add one row on the top of them.
So the $i$th row in $A$ or $B$ will be the $(i+1)$th row in $R$. The elements in $A$ now also appear in the first row of $R$. 
So the elements
in $A$ are appears in first row of $R$ and the rows where the row numbers are $Z$ 
combinations of $\{ 2, \dots, k+1\}$ in BT order,
 and the elements of $B$ appears in the
$Z + 1$ combinations of $\{ 2, \dots, k+1\}$ in BT order. So it is readily checked that the row numbers where 
the elements appear in $R$ are $Z +1$ 
combinations in $\{ 1, 2, \dots, k+1\}$ in BT order. 
The proof is complete.
\qed

\begin{theorem}\label{t.general+}
There exists an RPDA$(F, \ell{F\choose Z} - x, Z)$ for any integers $F \ge Z+1, x = 0, 1,  \dots, \left\lceil{Z+1\over F-Z}\right\rceil - 1$
and $\ell \ge 1$.
\end{theorem}
\p 
The results follows from Theorem \ref{t.general} and Lemmas \ref{l.times} and \ref{l.minus}.
\qed

\begin{corollary}\label{co.1}
For any integers $F \ge Z+1, x = 0, 1,  \dots,$ $ \left\lceil{Z+1\over F-Z}\right\rceil - 1$
and $\ell \ge 1$,
$$s(F, \ell{F\choose Z} - x, Z) = \frac{\ell{F\choose Z}(F-Z)}{Z+1}=\ell {F\choose Z+1}.$$
\end{corollary}

\begin{remark}\rm
One interesting characteristics of our construction of RPDAs is that an RPDA$(F, {F\choose Z}, Z)$ is
a sub-RPDA of an RPDA$(F+1, {F+1\choose Z}, Z)$. This property may be useful for the cases when 
the number of users increases or the database contents increase for the systems. 
\end{remark}

\section{Determine values of $s(F, K, Z)$}\label{s.s}

It is definitely a very interesting  problem to determine all of the values of $s(F,K, Z)$ for any combinations of  $F, K$ and $Z$.
In this section, we try to determine the values of $s(F,K,Z)$ for some special values of $F, K$ and $Z$ using  basic
combinatorial arguments.  

\subsection{$s(F,K,Z)$ for small $F$}

Suppose in a PDA one element appears $i$ times, then we can find a $i\times i$ sub-array which contains
 only
that element and other cells are empty by the definition of PDA. We say that the empty cells in that sub-array are
{\em shared} by that element. In a PDA, an empty cell may be shared by several elements. 

\begin{lemma}\label{l.share}
In a PDA$(F, K, Z)$, if $F \le K$, then an empty cell can be shared by at most $F - Z$ elements. If
$F> K$, then an empty cell can be shared by at most $K -1$ elements.  
\end{lemma}
\p
Suppose an empty cell $r_{i, j}$ is shared by an element $t$. Then $t$ must appear somewhere in the $i$th row
and somewhere  in the $j$th column. The conclusion follows.
\qed

First we consider the simplest cases when $Z = 1$.

\begin{theorem}\label{t.z=1}
Suppose $K = \ell F + i$, where $\ell \ge 0$ and $ 0\le i < F$. Then 
$$s(F, K, 1) = {\ell F(F-1)\over 2} + {i(i-1)\over 2} + i(F-i).$$
\end{theorem}
\p
For $i < F$, we can first construct a PDA$(F, i, 1)$ on an $i\times i$ array using Construction \ref{c.1}.
Then we add $F - i$ rows with different elements. It is easy to see that the $S$ set  of the PDA$(F,i,1)$ is the smallest.
In fact, we can see that each empty cell of the PDA is shared by $i -1$ different elements which is the maximum
according to Lemma \ref{l.share}.

For general cases ($\ell \ge 1$), consider the empty cells in the PDA. An empty cell will be shared
by at most $F-1$ elements by Lemma \ref{l.share}. If an empty cell is shared by $F-1$ elements, 
then these elements must appear in the same row and same column with that empty cell.
 We can mark these elements  together
with that empty cell. We could have marked at most 
${\ell F(F-1)\over 2}$ elements, each of them appears twice in the array. 
In this way, each row will have at least $i$ unmarked  cells left. These cells are either empty
cell or filled with an element that shared an empty cell with at most $i-1$ elements. And there are at
least $i (F-i)$ elements which do not share any empty cells. For the left elements,
only $i-1$ of them may appear twice in the array. But all of the other elements appears only once in
the array. Therefore $|S| \ge {\ell F(F-1)\over 2} + {i(i-1)\over 2} + i(F-i).$
 
We can concatenate $\ell$ RPDA$(F, F, 1)$s together,  then add the PDA$(F, i,1)$ to obtain the 
required PDA$(F, K, 1)$.
The conclusion follows from Corollary \ref{co.1}.
\qed

After we determined the cases of $Z = 1$ and  $Z = F-1$ (Lemma \ref{l.simple}), the next cases of $s(F, K,Z)$ 
we will consider are the cases of $F=4, Z =2$.

The bound giving in Theorem \ref{t.lowbound} depends on the maximum number of appearance of each of the 
elements in $S$. However, in many cased, the number of appearance, we call it the {\em frequency} of the element,
 cannot reach $Z+1$ if we want to keep the size of $S$ as small as possible.   
 
 \begin{lemma}\label{l.appear}
 If one element appears $Z+1$ times in an $S$-PDA$(F,K,Z)$, then $|S| \ge (F- Z - 1)(Z + 1) + 1$. 
 \end{lemma}
 \p
Suppose the element 1 appears $Z+1$ times in the PDA. Then the PDA contains a $(Z+1)\times (Z+1)$
sub-array $A$ which likes an ``identity matrix'' of order $Z + 1$.
  The  $(F - (Z + 1))\times (Z+1)$ sub-array   which consists
of the other rows outside 
$A$ must contain different elements. The conclusion follows.
\qed

\begin{example}\label{e.553}\rm
According to Lemma \ref{l.appear}, if a PDA$(5,5,3)$ has one element
appearing 4 times, then $|S| \ge 5$ and  if a PDA$(6,4,3)$ has one element appearing 4 times, then $|S| \ge
9$. In the following arrays, each element appears 3 times. So $s(5,5,3) = 4$ and $s(6,4,3) = 4$. 

\[
\begin{array}{|c|c|c|c|c|}
\hline
1&&3&&\\
\hline
2&&&&3\\
\hline
&1&&3&\\
\hline
&2&&&4\\
\hline
&&1&2&\\
\hline
\end{array}
\hskip 12pt
\begin{array}{|c|c|c|c|}
\hline
&&1&4\\
\hline
&1&&3\\
\hline
1&&&2\\
\hline
&4&3&\\
\hline
4&&2&\\
\hline
3&2&&\\
\hline
\end{array}
\]
\end{example}

\

\begin{lemma}\label{l.small}
$s(4, 1, 2)= 2;s(4,2,2)= 2; s(4,3,2) = 3; s(4,4,2) =4.$
\end{lemma}
\p
For $K = 1$ or $2$, the proof is straightforward.

 For $K = 3$, if one element appears 3 times, then $|S| \ge 4$ by Lemma \ref{l.appear}. 
In the  PAD$(4,3,2)$ showed in Figure \ref{f.small}, each element appears twice in the array.
So the $|S| = 3$ is the smallest.  

Similarly, for $K = 4$, if  there is an element appearing 3 times, then $|S| \ge 5$ by Lemma \ref{l.appear}.
  The Figure \ref{f.small} gives a PDA$(4,4, 2)$, in which each element appears 2 times in the array.
  So $s(4, 4, 2) = 4$.
\qed

\begin{figure}[h]
\begin{center}
$
\begin{array}{|c|c|}
\hline
1&\\
\hline
2&\\
\hline
&1\\
\hline
&2\\
\hline
\end{array}
$
\hskip 12pt
$
\begin{array}{|c|c|c|}
\hline
1&&3\\
\hline
2&&\\
\hline
&3&\\
\hline
&1&2\\
\hline
\end{array}
$
\hskip 12pt
$
\begin{array}{|c|c|c|c|}
\hline
1&&3&\\
\hline
2&&4&\\
\hline
&1&&3\\
\hline
&2&&4\\
\hline
\end{array}
$
\caption{$s(4,K,2)$ for small $K$}\label{f.small}
\end{center}
\end{figure}

In what follows, let $a_i$ be the number of elements that appear $i$ times in the PDA, where 
$i = 1, 2,\dots, Z+1$. Then 

\begin{eqnarray}
\sum_{i = 1}^{Z+ 1} i a_i &=& K( F - Z) \\
|S| &= &\sum_{i = 1}^{Z+ 1} a_i .
\end{eqnarray}

We want to minimize the value of $|S|$ under the constraints of PAD and the parameters. Basically,
we first want the value of $a_{Z+1}$ as large as possible. Then consider the values of $a_Z, a_{Z -1},
\dots $ etc. 

 \begin{theorem}\label{t.s(4,k,2)}
$$
s(4, K, 2) = \left\{ 
\begin{array}{ll}
\left\lceil{2K\over 3}\right\rceil& \mbox{if $K \equiv 0, 2$ or $5 \pmod6$}\\
\\
\left\lceil{2K\over 3}\right\rceil + 1& \mbox{otherwise}
\end{array}\right.
$$
\end{theorem}
\p
$s(4, K, 2) \ge \left\lceil{2K\over 3}\right\rceil$ by Theorem \ref{t.lowbound}.
For $K = 6\ell$ or $6\ell -1$, where $\ell \ge 1$, the results come from Lemma \ref{l.462}. For
$K = 6\ell +2$, we can construct an RPDA of $K = 6\ell$ and concatenate a PDA$(4, 2, 2)$ from
Lemma \ref{l.small} to get the required PDA.

For $K = 6\ell + i, i = 1, 3$ or $4, \ell \ge 1$, each empty cell in a PDA$(4, K, 2)$ shared by at
most 2 elements by Lemma \ref{l.share}. There are total $2(6\ell + i)$ empty cells in the PDA.
So there are possible $4(6\ell + i)$ shared spaces. If one element appear 3 times in the array, then it
needs $6$ shared spaces. And the two empty cells of the column
where the element is must be shared by that element. Suppose there are at least $4\ell$ elements, 
each of them appears 3 times.
Then there are  at most $2i$ empty cells left  (each of them may be shared by at most 2 elements) 
which are not shared for any of the $4\ell$ elements. Note that if an element appears 3 times in
the array then the two empty cells in the column of that element both shared by that element. So
 the $2i$
empty cells must be distributed in at most $i$ columns unless there are less than
$4\ell$ elements that appear 3 times in the array, or there are  less than $2i$ empty cells left.
 Then from Lemma \ref{l.small} we see that
there should be at least $i$ elements to fill these columns, each of the elements appears at most twice in the array.
 It is easy to know that if  $a_3 < 4\ell$, then the size of $S$ will be bigger. Now from Lemmas
\ref{l.462} and \ref{l.small}, we can obtain the required PDAs and complete the proof.
\qed 

It is easy to see that $s(F, K, Z) \le s(F, K+1, Z)$. We will use this fact in the following proofs.

\begin{lemma}\label{l.small5-3}
$s(5, 1, 3)=s(5,2,3)= 2; s(5,3,3)=s(5,4,3) = 3; s(5,5,3)= s(5,6,3) =4, s(5,7,3) = s(5,8,3) = s(5,9,3) = s(5,10,3) = 5.$
\end{lemma}
\p
It is easy to see the result of $s(5, K,3)$ for $K = 1, 2$ are true. For $K = 3$, two different elements cannot fill the
three columns, so the PDA$(5,3,3)$ in Figure \ref{f.small5-3} is optimal. Since $s(5,3,3)= 3,$
the PDA$(5,4,3)$ in Figure \ref{f.small5-3} is optimal. $s(5,5,3) = 4$ from Example \ref{e.553}
                                                                                                                                                                                                                                                  and thus the PDA$(5,6,3)$ in that Figure \ref{f.small5-3} is also optimal. In the PDA$(5,6,3)$ each of the 4 elements appears
3 times. Therefore $s(5,7,3) \ge 5$ no matter whether there is an element appears 4 times in the array (see Lemma \ref{l.appear}).
 From Corollary
\ref{co.1}, we have $s(5, 10, 3) = 5$. So $s(5,7,3) = s(5,8,3) = s(5,9,3) = s(5,10,3) = 5.$
\qed

\begin{figure}[h]
\begin{center}
$
\begin{array}{|c|c|}
\hline
1&\\
\hline
2&\\
\hline
&1\\
\hline
&2\\
\hline
&\\
\hline
\end{array}
$
\hskip 12pt
$
\begin{array}{|c|c|c|}
\hline
1&&3\\
\hline
2&&\\
\hline
&1&\\
\hline
&2&\\
\hline
&&1\\
\hline
\end{array}
$
\hskip 12pt
$
\begin{array}{|c|c|c|c|}
\hline
1&&3&\\
\hline
2&&&\\
\hline
&1&&3\\
\hline
&2&&\\
\hline
&&1&2\\
\hline
\end{array}
$
\hskip 12pt
$
\begin{array}{|c|c|c|c|c|c|}
\hline
1&&&3&4&\\
\hline
&1&&2&&4\\
\hline
&&1&&2&3\\
\hline
2&3&&&&\\
\hline
&&4&&&\\
\hline
\end{array}
$
\caption{$s(5,K,3)$ for small $K$}\label{f.small5-3}
\end{center}
\end{figure}

Using Lemma \ref{l.small5-3}, we can prove the following Theorem \ref{t.s(5,k,3)}. The proof  is very similar to
that of Theorem \ref{t.s(4,k,2)}, so we omit the details here.

\begin{theorem}\label{t.s(5,k,3)}
$$
s(5, K, 3) = \left\{ 
\begin{array}{ll}
\left\lceil{K\over 2}\right\rceil& \mbox{if $K \equiv 0$ or $9 \pmod{10}$}\\
\\
\left\lceil{K\over 2}\right\rceil + 1& \mbox{otherwise}
\end{array}\right.
$$
\end{theorem}
\qed

Using the similar method, we can prove the following.

\begin{lemma}\label{l.s(5,K,2)}
$s(5,3,2) = 5, s(5,4,2) =6, s(5,5,2)= 7, s(5,6,2) =8, s(5,7,2) = s(5,8,2)=s(5,9,2) =s((5,10,2)=10.$
\end{lemma}
\p
Since the proofs are very similar to that of the above lemmas, we omit the details but just provide the 
optimal PDAs in Figure \ref{f.5K2}. $s(5,5,2)$ will be proved in Lemma \ref{l.FF2} and $s(5,10, 2)$ are from 
Theorem \ref{t.general}.
\qed

\begin{figure}[h]
\begin{center}
$
\begin{array}{|c|c|c|}
\hline
1&&5\\
\hline
2&&4\\
\hline
3&4&\\
\hline
&1&3\\
\hline
&2&\\
\hline
\end{array}
$
\hskip 12pt
$
\begin{array}{|c|c|c|c|}
\hline
1&&4&5\\
\hline
2&4&&6\\
\hline
3&5&6&\\
\hline
&1&&3\\
\hline
&&2&\\
\hline
\end{array}
$
\hskip 12pt
$
\begin{array}{|c|c|c|c|c|c|}
\hline
&&1&3&7&6\\
\hline
1&&&4&5&8\\
\hline
&1&&&2&\\
\hline
3&4&5&&&2\\
\hline
6&7&8&2&&\\
\hline
\end{array}
$
\hskip 12pt
$
\begin{array}{|c|c|c|c|c|c|c|}
\hline
&&1&2&9&&5\\
\hline
1&&&4&6&3&7\\
\hline
&1&&8&&10&\\
\hline
2&3&4&&8&&10\\
\hline
5&6&7&&&9&\\
\hline
\end{array}
$
\caption{$s(5,K,2)$ for small $K$}\label{f.5K2}
\end{center}
\end{figure}

We then have the following results.

\begin{theorem}\label{t.s(5,k,2)}
$$
s(5, K, 2) = \left\{ 
\begin{array}{ll}
K& \mbox{if $K \equiv 0 \pmod{10}$}\\
K + 1& \mbox{if $K \equiv 9 \pmod{10}$}\\ 
K + 2& \mbox{otherwise}
\end{array}\right.
$$
\end{theorem}
\p We omit the details of the similar proofs. 
\qed

\begin{remark}\rm
Basically, to determine $s(F, K, Z)$ for fixed $F$ and $Z$, we need to construct optimal
PDA$(F, K, Z)$ for $K < {F\choose Z}$.
\end{remark}

\subsection{$s(F,K, Z) $ for large $F$}

Now we consider some examples of PDAs with large $F$ while $K$ and $Z$ are fixed.
First we note that the transpose of RPDA$(F,{F\choose Z}, Z)$ is a PDA$({F\choose Z}, F, {K-1\choose Z - 1})$.

\begin{lemma}\label{l.transpose}
The  transpose of the RPDA$(F,{F\choose Z}, Z)$ constructed in Theorem \ref{t.general} 
is a PDA$\left({F\choose Z}, F, {F -1\choose Z - 1}\right)$.
\end{lemma}
\p
To prove the result, we just need to show that each row of the RPDA$(F,{F\choose Z}, Z)$ 
constructed in Theorem \ref{t.general} has ${F -1\choose Z-1}$ empty cells. We prove that by
induction. In each row of the  RPDA$(Z + 1, {Z+1\choose Z}, Z)$, there are ${Z\choose Z-1}$ empty cells.
Assume that each row of the  RPDA$(F-1 ,{F -1\choose Z}, Z)$ has ${F-2\choose Z - 1}$ empty
cells and each row of the RPDA$(F-1, {F-1\choose Z-1}, Z-1)$ has ${F - 2\choose Z-2}$ empty cells.
 The first row of the RPDA$(F, {F\choose Z}, Z)$ constructed from Theorem \ref{t.general} has $F-1 \choose Z-1$
 empty cells. For each of the other row,  the number of empty cells is
 $${F - 2\choose Z-2}+ {F-2\choose Z - 1} = {F-1\choose Z-1}.$$
The proof is completed.
\qed

We consider a PDA$({K\choose t}, K, {K-1\choose t - 1})$ where $K > t > 1$ (the case of $t=1$ is trivial).

\cite{MN} provided a construction of PDA$\left({K\choose t}, K, {K -1\choose t - 1}\right)$ with $|S| = 
{K\choose t+1}$ which has been approved to be optimal according to
 the transmission rate. The transmission rate optimal is stronger then the PDA optimal. So we have the following result.

\begin{theorem}\label{t.largeF}
$s\left({K\choose t}, K, {K-1\choose t - 1}\right) = \frac{{K\choose t}(K - t)}{t+1}$, where  $K > t > 1$.
\end{theorem}
\p
The PDA in Lemma \ref{l.transpose} is a PDA$\left({K\choose t}, K, {K-1\choose t - 1}\right)$ with 
$|S| =  
\frac{{K\choose t}(K - t)}{t+1} = {K\choose t+1}$.
\qed

Next we consider the cases where the values $K$ and $Z$ are fixed, but $F$ will be arbitrary large.

We consider the cases for any $F$ with $K =4$ and $Z = 3$.

\begin{figure}[h]
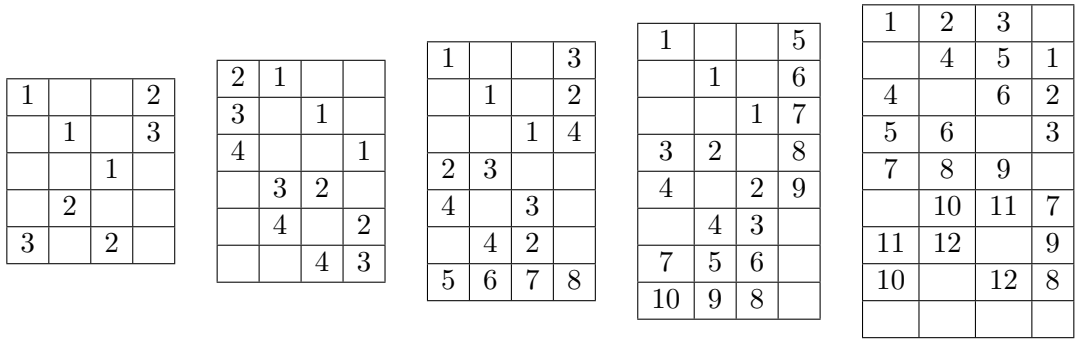

\begin{center}
$
\begin{array}{|c|c|c|c|}
\hline
1&&&2\\
\hline
&1&&3\\
\hline
&&1&\\
\hline
&2&&\\
\hline
3&&2&\\
\hline
\end{array}
$
\hskip 12pt
$
\begin{array}{|c|c|c|c|}
\hline
2&1&&\\
\hline
3&&1&\\
\hline
4&&&1\\
\hline
&3&2&\\
\hline
&4&&2\\
\hline
&&4&3\\
\hline
\end{array}
$
\hskip 12pt
$
\begin{array}{|c|c|c|c|}
\hline
1&&&3\\
\hline
&1&&2\\
\hline
&&1&4\\
\hline
2&3&&\\
\hline
4&&3&\\
\hline
&4&2&\\
\hline
5&6&7&8\\
\hline
\end{array}
$
\hskip 12pt
$
\begin{array}{|c|c|c|c|}
\hline
1&&&5\\
\hline
&1&&6\\
\hline
&&1&7\\
\hline
3&2&&8\\
\hline
4&&2&9\\
\hline
&4&3&\\
\hline
7&5&6&\\
\hline
10&9&8&\\
\hline
\end{array}
$
\hskip 12pt
$
\begin{array}{|c|c|c|c|}
\hline
1&2&3&\\
\hline
&4&5&1\\
\hline
4&&6&2\\
\hline
5&6&&3\\
\hline
7&8&9&\\
\hline
&10&11&7\\
\hline
11&12&&9\\
\hline
10&&12&8\\
\hline
&&&\\
\hline
\end{array}
$
\caption{$s(F,4,3)$ for $F = 5,6,7,8, 9$}\label{f.largeF}
\end{center}
\end{figure}

\begin{lemma}\label{l.F4-9}
$s(4,4,3) = 1, s(5,4,3) = 3, s(6,4,3) = 4, s(7,4,3) = 8, s(8,4,3) = 10$ and $ s(9,4,3) = 12$.
\end{lemma}
\p
From Lemma \ref{l.appear}, if an element in  the PDA$(F, 4,3)$ appears 4 times then $|S| \ge 4(  F - 4) + 1$. 
So the PDAs in Figure \ref{f.largeF} for $F = 5$ and $6$ are optimal ($F= 6$ can also be from Theorem \ref{t.largeF}). 
For the rest of the cases,
we are going to prove that each of the PDAs in Figure
\ref{f.largeF} is optimal. From Lemma \ref{l.appear}, we assume that no element appear 4 times in these PDAs.


\begin{itemize}
\item $F = 7 : a_1 + 2a_2 + 3a_3 = 16$. Since there are 16 cells fill with elements, there are at least two 
rows containing 3 elements or one row containing 4 elements.

 If there are two rows containing 3 elements, then there are at least 5 different elements in these two rows.
  Suppose  there are two rows, each of them contains 3 elements and there are 5 different elements,
 then the two rows will be something similar to the follows  
 $$
 \begin{array}{|c|c|c|c|}
 \hline
 1&&2&3\\
 \hline
 &1&4&5\\
 \hline
 \end{array}
 $$
 Then each of the first two columns have 2 empty cells left and thus there are at most one element which
 may appear 3 times (the elements $1,\dots, 5$ only can appear at most 2 times each). 
  So $a_3 \le 1$ which
 implies $a_1= 3, a_2 = 5$ or $a_1 = 1, a_2 = 6$ (we assume $a_3 = 1$. If $a_3 = 0$, then the size of $S$
 will be larger.) So in this case, $|S| \ge 8$.  
 
 If there is one row containing 4 elements, then these elements only can appear once in the PDA and
 $a_1 \ge 4$. The best case is that each of the other elements
 appears 3 times which implies $a_3 = 4, a_2 = 0$. Therefore $|S| = 8$. 
 
 \item $F = 8: a_1 + 2a_2 + 3a_3 = 20$. 
 
 If there are 4 rows having 3 elements and other 4 rows having 2 elements. Then the first 4 rows contain at least
 6 different elements, since each of the elements appear in the array at most 2 times. So $a_2 \ge 6$ and the best
 solution in this case is $(a_1, a_2, a_3 ) = (2, 6, 2)$ or $(1, 8, 1)$. So $|S| \ge 10$.
 
 If there are 2 rows having 3 elements, one row having 4 elements and 5 rows having 2 elements. Then we have
 $a_2 \ge 5$ (see the proof in case of $F = 7$), $ a_1 \ge 4$. And $a_3 > 0$ if $a_2 = 5$ and $a_1 = 4$. Again
 $|S| \ge 10$.
 
 If there are 2 rows having 4 elements, then $a_1 \ge 8$. Since
 $s(6, 4, 3) = 4$, we have $|S| \ge 12$.
 
 \item
 $F = 9: a_1 + 2a_2 + 3a_3 = 24$.
 
 If one element appears 3 times, then there are 21 nonempty cells left. We have $a_2 + a_1 \ge 11$ and $|S| \ge 12.$
 
 If  two elements appear 3 times, then these two elements appear in 5 rows (since $Z = 3$, they cannot appear in 6 rows
 but they also cannot just  appear in 4 rows). The 5 rows will be something similar to the following.
 $$
\begin{array}{|c|c|c|c|}
\hline
1&&&2\\
\hline
&1&&*\\
\hline
&&1&*\\
\hline
*&2&&*\\
\hline
*&&2&*\\
\hline
\end{array}
$$
 Since each of three columns has already 3 empty cells, the rest 4 rows of these columns have no
 empty cells. The elements in these cells must be different and thus $|S| \ge 14$. From the above array,
 we can also see that it is impossible to have 3 elements, each of them appear 3 times.
 
 If each of the elements appears 2 times, then $|S| = 12$. 
 
 If one element appears once, then other 23 nonempty cells contain at least 12 different elements and 
 $|S| \ge 13$.
\end{itemize}

\begin{remark}\rm
The PDA$(9,4,3)$ in Figure \ref{f.largeF} is also a PDA$(8, 4, 2)$. This interesting case shows that
in some cases, increasing a little memory of the nodes may not improve the  network traffic of communication.
\end{remark}

\begin{lemma}\label{l.F10-12}
$s(10,4,3) = 14, s(11,4,3) = 17$ and $ s(12,4,3) = 18$.
\end{lemma}
\p
We are going to prove that the PDAs in Figure \ref{f.largeF2} are all optimal. In that figure, the first array is a 
PAD$(10, 4, 3)$. If we exchange the bottom two rows of the PDA$(10, 4, 3)$
with each of the other two arrays in the figure, we obtain the PDA$(F, 4,3)$
for $F = 11$ and $12$ respectively.
\begin{itemize}
\item
$F = 10: a_1 + 2a_2 + 3a_3 = 28$.

If at least one row contains 4 elements, then other 9 rows contain at least 12 different elements since $S(9,4,3) = 12$.
So $|S| \ge 16$.  

If no row contains 4 elements, then no element can appear 3 times in the array since there are at most
2 rows containing 2 elements.  Therefore each element appears at most 2 times in the array which implies
$|S| \ge 14$.
\item
$F = 11: a_1 + 2a_2 + 3a_3 = 32$.

If at least one row contains 4 elements, then other 10 rows contain at least 14 different elements since $S(10,4,3) = 14$.
So $|S| \ge 18$.  

If no row contains 4 elements, then no element can appear 3 times in the array since there are at most
1 rows containing 2 elements.  Therefore each element appears at most 2 times in the array.  In this case, there are
10 rows, each of them contains 3 elements, and another row contains 2 elements. Each empty cell in the 10 rows can be 
shared by at most 3 elements, and each empty cells in the 11th row can be shared by at most 2 elements. Therefore there are
at most 34 shares can be used. If an element appear 2 times, then it needs 2 shares. 
  Therefore $|S| \ge 17$.
\item
$F = 12: a_1 + 2a_2 + 3a_3 = 36$.

Use a similar argument, we can see that in the optimal array, each element will appear 2 times. Therefore $|S| \ge 18$.
We omitted the details of proof for this case.
\end{itemize}
 \qed
 
\begin{figure}[h]
\begin{center}
$
\begin{array}{|c|c|c|c|}
\hline
1&&2&4\\
\hline
&1&3&5\\
\hline
3&2&&6\\
\hline
5&4&6&\\
\hline
8&7&&12\\
\hline
9&&7&10\\
\hline
&9&8&11\\
\hline
11&10&12&\\
\hline
\hline
&14&&13\\
\hline
14&&13&\\
\hline
\end{array}
$
\hskip 12pt
$
\begin{array}{|c|c|c|c|}
\hline
13&14&&16\\
\hline
15&&14&17\\
\hline
&15&13&\\
\hline
\end{array}
$
\hskip 12pt
$
\begin{array}{|c|c|c|c|}
\hline
13&14&16&\\
\hline
17&15&&16\\
\hline
18&&15&14\\
\hline
&18&17&13\\
\hline
\end{array}
$
\caption{$s(F,4,3)$ for $F = 10, 11, 12$}\label{f.largeF2}
\end{center}
\end{figure}

\begin{theorem}
$s(F, 4, 3) = 4F - 30$ for any $F \ge 12$.
\end{theorem}
\p
When $F \ge 12$, there are at least $F-12 $ rows containing 4 elements. From Lemma \ref{l.F10-12} we know
that $s(12, 4, 3) = 18$. Therefore if $F-12$ rows contain 4 elements, then $|S| \ge 18+ 4(F- 12) =
4F - 30.$ We have determined $s(F, 4, 3)$ for $F < 12$. Therefore 
it is readily checked that if there are more than $F - 12$ rows containing 4 elements,
then $|S| > 4F- 30$. The conclusion follows.
\qed

\begin{remark}\rm
For the cases of $F$ relatively large than $K$, we need to determine the cases for $F \le KZ$. If 
$F > KZ$, then there are at least $F - KZ$ rows filled with different elements. So it is reasonable to always 
assume that $F \le KZ$.
\end{remark}

\subsection{Some more examples}

We give some examples of optimal PDAs in which $F = K$.

\begin{lemma}\label{l.3t}
$s(3t, 3t, 3t-2) = 3$, for all  $t \ge 1$.
\end{lemma}
\p In a PDA$(3t, 3t, 3t-2) $, one element appears at most $3t - 1$ times. Since there are $6t$ nonempty cells,
$|S| \ge 3$. To construct $3$-PDA$(3t, 3t, 3t-2)$, we first construct an optimal PDA$(3, 3, 1)$:
$$
\begin{array}{|c|c|c|}
\hline
&3&2\\
\hline
3&&1\\
\hline
2&1&\\
\hline
\end{array}
$$
Then we substitute each cell with a $t\times t$ sub-array.  If the cell contains an element, then
 same element is put at  the cells in diagonal of the sub-array, and
other cells of the sub-array are empty. If the original cell is empty then all cells of the $t\times t$ 
substituting sub-array are empty. 
\qed

\begin{example}\rm
An optimal  PDA$(6,6,4)$ from Lemma \ref{l.3t} is as follows.
$$
\begin{array}{|c|c|c|c|c|c|}
\hline
&&3&&2&\\
\hline
&&&3&&2\\
\hline
3&&&&1&\\
\hline
&3&&&&1\\
\hline
2&&1&&&\\
\hline
&2&&1&&\\
\hline
\end{array}
$$
\end{example}

\begin{remark}\rm
Using a similar method of the proof of Lemma \ref{l.3t}, we can construct a $6$-PDA$(4t, 4t, 4t -3)$ for any $t \ge 1$
from the PDA below.
$$
\begin{array}{|c|c|c|c|}
\hline
1&2&4&\\
\hline
5&3&&4\\
\hline
6&&3&2\\
\hline
&6&5&1\\
\hline
\end{array}
$$
 
However, when $t \ge 4$ the PDAs may not be optimal. 
\end{remark}

\begin{example}\rm
An optimal  PDA$(8,8,5)$ from a PDA$(4,4,1)$ using the above remark is as follows.
$$
\begin{array}{|c|c|c|c|c|c|c|c|}
\hline
1&&2&&4&&&\\
\hline
&1&&2&&4&&\\
\hline
5&&3&&&&4&\\
\hline
&5&&3&&&&4\\
\hline
6&&&&3&&2&\\
\hline
&6&&&&3&&2\\
\hline
&&6&&5&&1&\\
\hline
&&&6&&5&&1\\
\hline
\end{array}
$$

To prove its optimality, we consider the largest frequency of the elements. If one element appears 6 times, then
$|S| \ge 7$ by Lemma \ref{l.appear}. If one element appears 5 times, then there is a sub-PDA$(3,5,1)$ which uses
at least 6 different elements by Theorem \ref{t.z=1}, and therefore $|S| \ge 7$. Since each element appears 4 times in the above 
PDA, it is optimal.

\end{example}
\begin{remark}\rm
Using a similar method with $t = 3$, 
we can prove that $s(12,12,9) = 6$ from the fact that $s(5,7,3)= 5$ (Theorem \ref{t.s(5,k,3)}).
\end{remark}

\begin{lemma}\label{l.FF2}
$s(3,3,2) = 1, s(4,4,2) = 4, s(5,5,2) = 7$.
\end{lemma}
\p
$s(4,4,2) = 4$ comes form Theorem \ref{t.s(4,k,2)}. For a PDA$(5, 5, 2), a_1+2a_2+3a_3 = 15$.
Suppose one element appears 3 times, then the first 3 columns must contain 7 different elements
and it is impossible to have more than one element appearing 3 times. If no element appears 3 times,
than $|S| \ge 8$. Therefore the PDA$(5, 5, 2)$ in Figure \ref{f.FF2s} is optimal.
\qed
\begin{figure}[h]
\begin{center}
$
\begin{array}{|c|c|c|}
\hline
&&1\\
\hline
1&&\\
\hline
&1&\\
\hline
\end{array}
$
\hskip 12pt
$
\begin{array}{|c|c|c|c|}
\hline
&&1&2\\
\hline
1&&&3\\
\hline
&1&&\\
\hline
2&3&4&\\
\hline
\end{array}
$
\hskip 12pt
$
\begin{array}{|c|c|c|c|c|}
\hline
&&1&2&5\\
\hline
1&&&3&6\\
\hline
&1&&4&7\\
\hline
2&3&4&&\\
\hline
5&6&7&&\\
\hline
\end{array}
$
\end{center}
\caption{$s(F, F, 2)$ for $F = 3, 4, 5$}\label{f.FF2s}
\end{figure}

For general $F$ and $Z = 2$, we have the following result.

\begin{theorem}\label{t.FF2}
$s(F, F, 2) = 3F- 8 + s(F-3, F - 3,2)$, for $F \ge 6$.
\end{theorem}
\p
For $F < 6$, we have optimal PDA$(F, F, 2)$ in which all of the cells in the main diagonal are empty 
from Lemma \ref{l.FF2} (for $F = 3$ and $4$, it is easy to construct the required PDAs) and each of them
has at least one element appears 3 times and at most one element appears one time.
 For $F \ge 6$, we also assume that there is at least one element appears 3 times in the PDA, which
are at the top left  $3\times 3$ sub-array in the PDA (see the figure below). 
\begin{center}
 \includegraphics[width=4.2 cm]{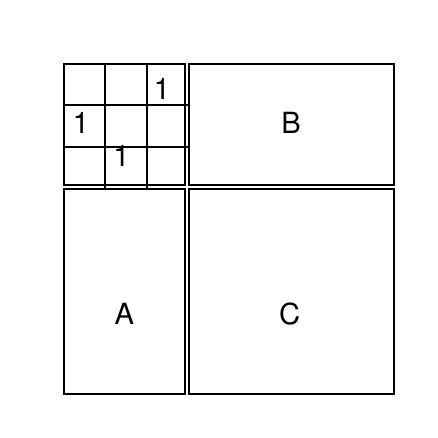}
\end{center}
Then
there are other $3(F - 3)$ different elements in the first 3 columns (the sub-array A in the figure). Denote these elements as a 
set $S_1$. Any element in $S_1$ cannot appear 
twice in the 
bottom $F-3$ rows according to the definition of PDA because there is no empty cells in A. We can prove that these elements should not appear
3 times in the PDA if the PDA is optimal. If otherwise 
there is $t\in S_1$ which appears 3 times, then $t$ will appear 2 times in sup-array B and there are at least
2 empty cells in B. In that case,
there are at least 3 elements in $S_1$, each of them only appears once in the PDA. Since all of the elements in sub-array C are not in $S_1$, $a_1$ will 
increase 3 while $a_3$ increases 1 and $a_2$ decreases 3. So in an optimal PDA,   each element in $S_1$ shall appear 2 times. Since we let the cells in the
main diagonal of the $3\times 3$ top right sub-array be empty, we can just let B be the transpose of A if the cells in the main diagonal of C
are also all empty. Now we can see that the sub-array C is a 
PDA$(F-3, F-3, 2)$ since the elements in C do not appear out side C, and each column of C has two empty cells. In our optimal PDA$(F, F, 2)$ for
$F < 6$, the cells in the main diagonal are empty. By induction, we can let all of the PDA$(F, F, 2)$ have empty cells in their main diagonal.
So the smallest size of $|S| = 3(F-3) 
+ 1 + s( F-3, F-3, 2)$. The above proof actually gives a recursive construction. Note that in the construction, at most one element appears once in the PDA,
but other elements appear at least 2 times each. 

We can see that if no element appears three times in the array, then $|S| \ge \left\lceil {(F-2)F\over 2}\right\rceil$. It is not difficult to prove that
$ \left\lceil {(F-2)F\over 2}\right\rceil > 3F - 8 
  + s( F-3, F-3, 2)$ for $F \ge 6$. For $F= 3, 4, 5, 6$, it can be direct verified. For $F > 6$, by induction
 \begin{eqnarray*}
 3F - 8 + s( F-3, F-3, 2)& < & 3F-8  + {(F-3-2)(F-3)\over 2}\\
 &=& {(F-2)F -1\over 2}. 
 \end{eqnarray*} 
 The proof completed.
\qed

Next we give the examples of $F = K = 6 $ and $7$ for all possible values of $Z$.

\begin{lemma}
$s(6,6,1) = 15, s(6,6,2)= 11, s(6,6,3)= 6, s(6,6,4)= 3, s(6,6,5) = 1$
\end{lemma}
\p
The cases of $Z = 1$ and $5$ are from Lemmas \ref{l.simple} and \ref{l.z=1}. 
The case of  $Z =2$ is from Theorem \ref{t.FF2}. The case of $Z = 4$ is from 
Lemma \ref{l.3t}.   

When 
$Z = 3,  a_1 + 2a_2 + 3a_3 + 4a_4 = 18.$
If $a_4 \ge 1$, then $|S| \ge 9$ by Lemma \ref{l.appear}. Let $a_4= 0, a_3 = 6$. 
If one element appears 3 times in the array, then it shares 6 empty cells. Therefore 
the array needs 36 shares of empty cells. There are total 18 empty cells each of them
only can provide 3 shares. Therefore there are at most  we obtain the optimal
PDA in   Figure \ref{f.F=K}.
\qed

\begin{figure}[h]
\begin{center}
$
\begin{array}{|c|c|c|c|c|c|}
\hline
&&1&2&5&8\\
\hline
1&&&3&6&9\\
\hline
&1&&4&7&10\\
\hline
2&3&4&&11&\\
\hline
5&6&7&&&11\\
\hline
8&9&10&11&&\\
\hline
\end{array}
$
\hskip 12pt
$
\begin{array}{|c|c|c|c|c|c|}
\hline
1&&&2&3&\\
\hline
&1&&4&&3\\
\hline
&&1&&4&6\\
\hline
6&&2&&5&\\
\hline
4&2&&&&5\\
\hline
&6&3&5&&\\
\hline
\end{array}
$
\caption{$s(6,6,Z)$ for $Z = 2,3$}\label{f.F=K}
\end{center}
\end{figure}

For $F = K = 7$, we have the following result. We omit the proved cases of $Z = 1, 2 $ and $6$ in the following lemma.
\begin{lemma}\label{l.F=K7}
$ s(7,7,3)= 10, s(7,7,4) =6, s(7,7,5)= 4.$
\end{lemma}
\p
 We list the optimal
PDAs in Figure \ref{f.F=K7} and give the proofs below.

\begin{itemize}
\item
$Z = 3: a_1 + 2a_2 + 3a_3 + 4a_4 = 28.$

If $a_4 \ge 1$, then $|S| \ge 13$ by Lemma \ref{l.appear}. Let $a_4= 0$ then $a_3 \le 9.$ We have solutions
$(a_1, a_2, a_3, a_4) = (1, 0,9,0)$ or $(0,2, 8,0)$. If $a_3 < 8$, then $|S| > 10$.

\item
$Z = 4: a_1 + 2a_2 + 3a_3 + 4a_4+ 5a_5 = 21.$

If $a_5 \ge 1$, then $|S| \ge 11$ by Lemma \ref{l.appear}. Let $a_5= 0,$ then  $a_4 \le 5$ and $|S| \ge 6$. 
So the PDA$(7,7, 4)$ in Figure \ref{f.F=K7} is optimal.

\item 
$Z = 5: a_1 + 2a_2 + 3a_3 + 4a_4+ 5a_5 + 6a_6= 14.$

If $a_6 \ge 1$, then $|S| \ge 7$ by Lemma \ref{l.appear}.  If one element appears 5 times 
in the PDA, then there cannot be another element appear 5 times because only 2 rows and two columns left to 
contain other elements. So the best solutions are $(a_1,a_2,a_3,a_4,a_5,a_6) = (1, 0, 0, 2, 1,0)$ or $(0,0,3,0,1,0)$. 
If $a_6 =a_5= 0$, then $|S| \ge 4$ since there are 14 nonempty cells. 
\end{itemize}
\qed

\begin{figure}[h]
\begin{center}
$
\begin{array}{|c|c|c|c|c|c|c|}
\hline
1&&&2&6&7&\\
\hline
&1&&3&5&&7\\
\hline
&&1&4&&5&6\\
\hline
3&2&&&9&10&\\
\hline
4&&2&&8&&10\\
\hline
&4&3&&&8&9\\
\hline
5&6&7&8&&&\\
\hline
\end{array}
$
\hskip 11pt
$
\begin{array}{|c|c|c|c|c|c|c|}
\hline
1&&&&2&&4\\
\hline
&1&&&&2&6\\
\hline
& &1&&&5&3\\
\hline
&&&1&6&4&\\
\hline
5&3&2&&&&\\
\hline
6&4&&2&&&\\
\hline
&&4&3&5&&\\
\hline
\end{array}
$
\hskip 11 pt
$
\begin{array}{|c|c|c|c|c|c|c|}
\hline
1&&&&&&3\\
\hline
&1&&&&3&\\
\hline
& &1&&&&2\\
\hline
&&&1&&2&\\
\hline
&&&&1&&\\
\hline
2&&4&3&&&\\
\hline
&2&&&3&&\\
\hline
\end{array}
$
\caption{$s(7,7,Z)$ for $Z = 3,4,5$}\label{f.F=K7}
\end{center}
\end{figure}

\begin{example}\rm
In general, there are different distributions of the elements in optimal PDAs.
Figure \ref{f.more} displays an
 optimal PDA$(7,7,5)$  which are different from that in the proof of Lemma \ref{l.F=K7}.
\end{example}

\begin{figure}[h]
\begin{center}
$
\begin{array}{|c|c|c|c|c|c|c|}
\hline
1&&&&&&2\\
\hline
&1&&&&2&\\
\hline
&&1&&2&&\\
\hline
&&&1&3&&\\
\hline
&&3&2&&&\\
\hline
&3&&&&4&\\
\hline
3&&&&&&4\\
\hline
\end{array}
$
\caption{PDA$(7,7,5)$}\label{f.more}
\end{center}
\end{figure}

The method used in Lemma \ref{l.3t} can be used to prove the following construction.

\begin{lemma}\label{l.twice}
If there is an $S$-PDA$(F, K, Z)$, then there is an $S$-PDA$(2F, 2K, F+Z)$.
\end{lemma}
\p
Substitute each element in the PDA$(F, K, Z)$ with a $2\times 2$ sub-array using a method
similar to that in the proof of Lemma \ref{l.3t}. Now each column
has $2Z + (F-Z) = F + Z$ empty cells.
\qed

\begin{corollary}
If $\left\lceil {2K(F - Z)\over F+ Z + 1}\right\rceil = \left\lceil {K ( F - Z)\over Z+1}\right\rceil$
and there is an RPDA$(F, K, Z)$, then
 $$s(2F, 2K, F + Z)=  \left\lceil {K ( F - Z)\over Z+1}\right\rceil.$$
\end{corollary}
\p
Since $s(2F, 2K, F + Z) \ge \left\lceil {2K(F - Z)\over F+ Z + 1}\right\rceil$, the conclusion comes
from Lemma \ref{l.twice}.
\qed

\begin{example}\rm
From the above Corollary, $s(8, 12, 6) = 4$, since there is an RPDA$(4,6,2)$. And $s(10, 20,8) = 5$, since there is an
RPDA$(5, 10, 3)$.
\end{example}

\section{Conclusions}\label{s.conclusion}

 In this paper, we define optimization of PDA based coded caching schemes for given values of
  subpacketization, number of nodes,
 size of nodes' storage and size of data. This definition is weaker than the definition of information rate in \cite{MN},
 but it is useful under the constraints of real applications. 
 
 We determined the following optimal PDAs by combinatorial constructions:
 \begin{itemize}
 \item
 $s(F, K, F-1)$.
 \item
 $s(F, K, F-t)$, where $F \ge tK$.
  \item
 $s(F, K, 1)$.
  \item
 $s(F, K, F-1)$.
  \item
 $s(F, 2, Z)$.
 \item
 $s(F, F, 2)$.
  \item
 $s(3t, 3t, 3t-2)$ for $t \ge 1$.
  \item
 $s(F, \ell{F\choose Z}- x, Z)$, where $F \ge Z+1, x = 0, \dots, \lceil {Z+1\over F - Z}\rceil$ and $\ell \ge 1$.
  \item
 $s({K\choose t}, K, {K -1\choose t - 1})$, where $K > t > 1$.
 \end{itemize}
 We also determined the values of $s(F,K,Z)$ for some small cases:
 \begin{itemize}
 \item
 $s(4, K, 2), s(5, K, 2)$ and $s(5, K, 3)$ for any $K$. 
 \item
 $s(F, 4,3)$ for any $F$.  
 \item
 $s(6,6,Z)$ for $Z\le 5$.
 \item
 $s(7,7, Z)$ for $Z \le 6$.
 \end{itemize}
 Some combinatorial constructions are used for determining the optimal PDAs. However, for general $F, K$ and $Z$, we 
 need more powerful constructions for optimal PDAs 
 in the future. There are many interesting problems for optimal PDAs. We would like to 
 list some of them below, that are more closely related to combinatorics.
 \begin{itemize}
 \item
 Find some necessary conditions for the existence of RPDAs which are more strict than the condition in Lemma \ref{l.necessary}.
 Especially, do the conditions of $K$ in Theorem \ref{t.general+} necessary in general?
 \item
 Find a direct construction of RPDA$(F, {F\choose Z}, Z)$ in Theorem \ref{t.general} so that it is 
 not necessary to construct it recursively starting from the RPDA$(Z+1, Z+1, Z)$.
 \item
 Find general constructions for optimal PDA$(F, K, Z)$, where $F- Z$ is a fixed value.
 \item
 Find general constructions for optimal PDA$(F, F, Z)$, for $Z > 2$.
 \end{itemize}

\end{document}